\documentclass{osa-article}

\journal{ao}
\articletype{Research Article}
\usepackage{lineno}
\usepackage{soul}
\usepackage{siunitx}
\usepackage{comment}

\newcommand{\lt}{\ensuremath <}

\usepackage{gensymb}

\begin{document}

\title{Sidelobe Modeling and Mitigation for a Three Mirror Anastigmat Cosmic Microwave Background Telescope}

\author{Ian Gullett\authormark{1}, Bradford Benson\authormark{2,3,4}, Robert Besuner\authormark{5}, Richard Bihary\authormark{1}, John Carlstrom\authormark{2,3,6,7,8}, Nick Emerson\authormark{9}, Patricio A. Gallardo\authormark{2}, Jillian Gomez\authormark{1}, Cesiley L. King\authormark{10}, Jeff McMahon\authormark{2,3,4,6,7}, Jared L. May\authormark{10}, Johanna M. Nagy\authormark{10, 11}, Tyler Natoli\authormark{2,3}, Michael D. Niemack\authormark{12, 13}, Kate Okun\authormark{1}, Stephen Padin\authormark{14}, John E. Ruhl\authormark{1}, Edward J. Wollack\authormark{15}, Jeff Zivick\authormark{3}}

\address{\authormark{1}Physics Dept, Case Western Reserve University, Cleveland, OH, USA\\
\authormark{2}Kavli Institute for Cosmological Physics, University of Chicago,
5640 South Ellis Avenue, Chicago, IL, 60637, USA\\
\authormark{3}Department of Astronomy and Astrophysics, University of Chicago,
5640 South Ellis Avenue, Chicago, IL, 60637, USA\\
\authormark{4}Fermi National Accelerator Laboratory, Batavia, IL, USA\\
\authormark{5}Lawrence Berkeley National Laboratory, 1 Cyclotron Rd, Berkeley, CA 94720\\
\authormark{6}Department of Physics, University of Chicago, Chicago, IL, USA\\
\authormark{7}Enrico Fermi Institute, University of Chicago, Chicago, IL, USA\\
\authormark{8}High-Energy Physics Division, Argonne National Laboratory\\
\authormark{9}Steward Observatory, The University of Arizona, Tucson, Arizona, USA\\
\authormark{10}Department of Physics,  Washington University in St. Louis, St. Louis, MO, USA\\
\authormark{11}McDonnell Center for the Space Sciences, Washington University in St. Louis, St. Louis, MO, USA\\
\authormark{12}Department of Physics, Cornell University, Ithaca, NY, USA\\
\authormark{13}Department of Astronomy, Cornell University, Ithaca, NY, USA\\
\authormark{14}California Institute of Technology, Pasadena, CA, USA\\
\authormark{15}NASA Goddard Space Flight Center, 8800 Greenbelt Rd, Greenbelt, MD, USA}

\email{\authormark{*}ixg90@case.edu} %% email address is required



\begin{abstract}
Telescopes measuring cosmic microwave background (CMB) polarization on large angular scales require exquisite control of systematic errors to ensure the fidelity of the cosmological results.  In particular, far-sidelobe contamination from wide angle scattering is a potentially prominent source of systematic error for large aperture microwave telescopes.  Here we describe and demonstrate a ray-tracing-based modeling technique to predict far sidelobes for a Three Mirror Anastigmat (TMA) telescope designed to observe the CMB from the South Pole.  Those sidelobes are produced by light scattered in the receiver optics subsequently interacting with the walls of the surrounding telescope enclosure.  After comparing simulated sidelobe maps and angular power spectra for different enclosure wall treatments, we propose a highly scattering surface that would provide more than an order of magnitude reduction in the degree-scale far-sidelobe contrast compared to a typical reflective surface.  We conclude by discussing the fabrication of a prototype scattering wall panel and presenting measurements of its angular scattering profile. 
\end{abstract}

%Link to slides for LAT WG presentation: https://docs.google.com/presentation/d/1bySZ4qHcG8ez9MtXrp9weW23VkKsluBp4MS_3W_Xvbc/edit?usp=sharing

%%%%%%%%%%%%%%%%%%%%%%%%%%  body  %%%%%%%%%%%%%%%%%%%%%%%%%%
\section{Introduction}
Measurements of the cosmic microwave background (CMB) temperature and polarization anisotropies contain a wealth of information about the composition and evolution of the Universe.  One major goal of upcoming CMB instruments is to use measurements of the odd-parity CMB $B$-mode polarization to search for evidence of the primordial gravitational waves predicted by models of cosmic inflation. As current instruments produce stronger constraints on the amplitude of primordial gravitational wave signals (e.g.\cite{ BK_r, Spider_r, Planck_r, SPT_r}), future experiments will need to also precisely measure the gravitational lensing signal in CMB polarization with sufficient precision to separate it from the primordial component. Future generation experiments such as CMB-S4 \cite{S4_ref_design} are being designed with more than an order of magnitude more sensitivity to the primordial component than the upper limits of current experiments \cite{S4_r_forecast}. In addition to improving raw sensitivity, this measurement also requires correspondingly better control of systematic errors as well as observations in many frequency bands from $20-280$~GHz to separate Galactic foreground signals from the CMB.  

A new microwave telescope, designed for operation at the South Pole, has recently been described by \cite{padinTMA}.  This Three Mirror Anastigmat (TMA) offers a number of advantages for precise measurements of CMB polarization signals from both gravitational lensing and primordial gravitational waves, therefore a modified TMA has been chosen as the baseline South Pole Large Aperture Telescope by CMB-S4. The high optical throughput of the TMA accommodates the large number of detectors required to achieve the necessary sensitivity across all observing bands as described in \cite{PatoSPIE}. Additionally, the $5$~m diameter primary mirror provides the necessary angular resolution to measure the gravitational lensing signal. The use of monolithic mirrors eliminates sidelobes due to the gaps between panels in the segmented mirrors on previous large-aperture CMB telescopes. Furthermore, the telescope's ability to rotate about its boresight axis also mitigates some types of systematic polarization errors. 
Despite the added complexity of a design with three large mirrors and boresight rotation, the combination of these features makes this design well suited to measuring the large angular scale primordial “B-mode" signals from inflation, provided such systematic effects, in particular far sidelobe response, can be adequately controlled.
Power that spills over the mirrors and interacts with the interior of the enclosure surrounding the mirrors and receiver is an important source of such far sidelobes, and modeling and controlling this power is the main focus of this work .
For an off-axis telescope with no beam-blocking structures and no mirror panel gaps, diffraction effects are expected to be negligible at large angles \cite{SPT_George_etal}.

It is important to adequately control sidelobe pick-up in CMB telescopes because the complex and time-varying nature of such signals makes them difficult to remove from the data without suppressing cosmological information. 
Other large aperture CMB telescopes have conducted sidelobe modeling studies to understand the impact of this systematic effect on the cosmological results (e.g.\cite{ACT_sidelobes, ACT_sidelobes2, SO_optics_modeling, Adler_SPIE}). However, the unique design of this TMA and the specific goal of measuring large angular scale B-mode CMB polarization signals require a dedicated study.  Section \ref{sec: method} of this paper presents a ray-tracing-based sidelobe modeling technique applied to the TMA; the resulting sidelobe maps and their angular power spectra are presented in Section \ref{sec: results}. 

Our investigation considers several different types of surface treatments (reflective, scattering, and absorptive) on the interior telescope enclosure walls as well as the impact of combining data from detectors spread across the large focal plane. Ideally we would then convolve a given simulated sidelobe map with known-input signals from the sky and ground as the telescope scans across the observing field.  Unfortunately, the uncertain nature of the ground signal prohibits performing this convolution;  instead, we calculate the angular power spectrum of the sidelobe map as an indicator of its response as a function of angular scale, irrespective of the ground signal.  

We find that using scattering rather than reflective enclosure walls leads to significantly lower 
ground pick-up on the the relevant degree angular scales.
Similar highly scattering surfaces have been previously used in a variety of other optical applications at THz frequencies (e.g.\cite{Wang:21} \cite{Shirley:88}).  Section \ref{sec: proto_surfaces} presents measurements from a prototype scattering surface and discusses potential methods for fabrication.  Practical considerations for robustly mounting this material to the interior of the telescope enclosure are also discussed.  We find that the highly scattering enclosure offers a promising sidelobe mitigation strategy for this telescope design.

\section{Modeling Sidelobe Pick-Up with Ray Tracing Simulations}
 \label{sec: method}
\begin{figure}
 \begin{centering}
 \includegraphics[scale=0.6]{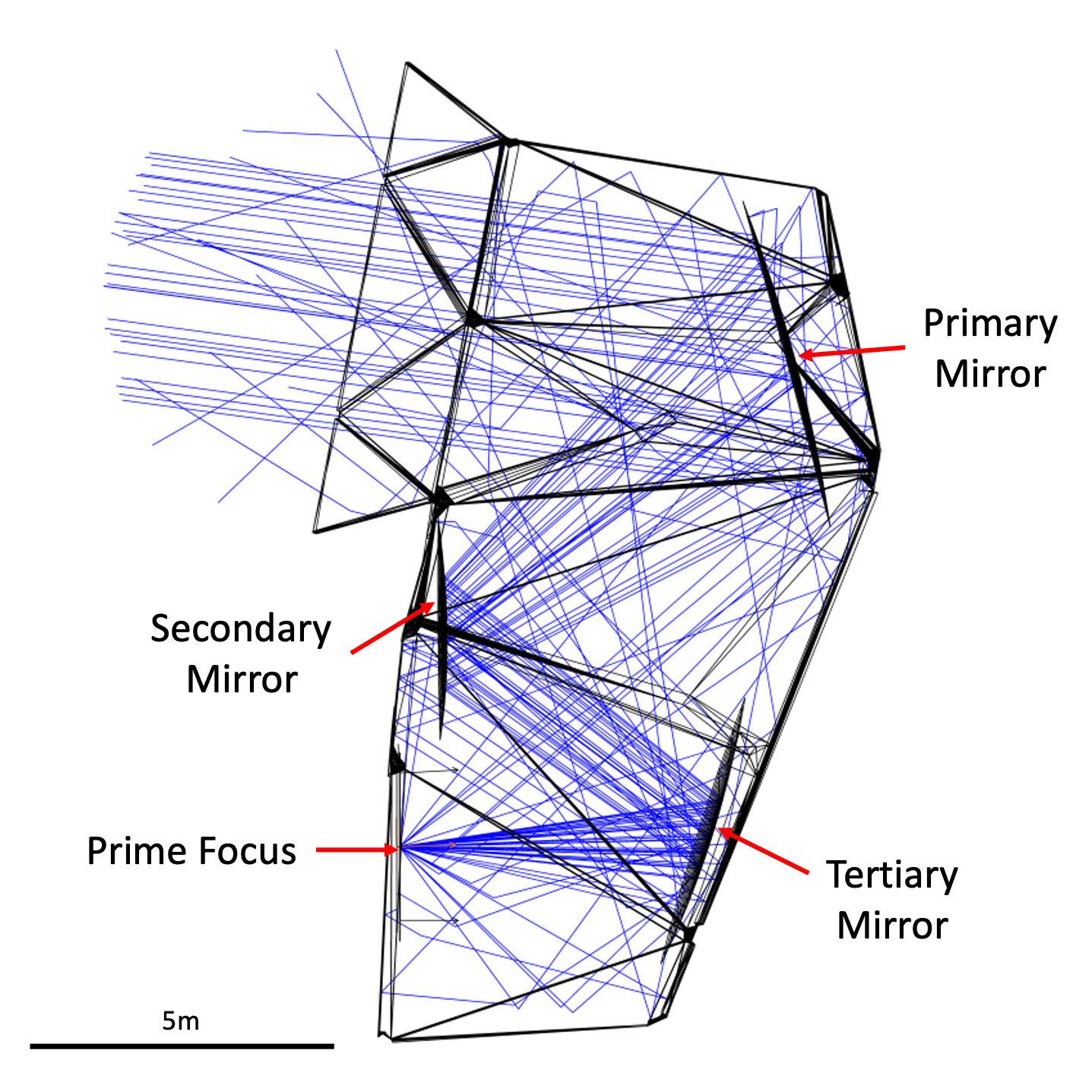}
 \par
 \end{centering}
 \caption{Diagram of the setup for the non-sequential ray tracing simulations.  In this time reversed model, ray bundles launched from the telescope focal plane propagate until they either reach the sky or are absorbed within the enclosure.  A fraction of the rays scatter to wide angles at the window of the cryogenic receiver, 50~mm in front of the focal plane. In this image, the enclosure walls are specularly reflective and the window scattering fraction has been increased to 50\% to illustrate a variety of possible ray paths. Note that some rays exit the telescope aperture at large angles relative to the main beam;  those large-angle rays form the sidelobes of concern in this paper.}
 \label{fig: setup}
 \end{figure}

To predict the sidelobe power expected from this telescope design with different enclosure-wall treatments, we perform ray tracing simulations to generate sidelobe maps.  Here we use the Zemax OpticStudio Software package\footnote{\url{https://www.zemax.com/pages/opticstudio}} in the non-sequential ray tracing mode and run the simulations in a time-reversed sense. Ray bundles are launched from various positions across the telescope's focal plane and propagated through the optics onto to the sky as shown in Figure \ref{fig: setup}.  A hemispherical detector surface skyward of the telescope records the intensity and angle of all the outgoing rays, allowing the reconstruction of the far-field beam pattern.

Based on measurements made by similar microwave telescopes, the wide angle beam sidelobes are likely to be dominated by rays that scatter in the receiver optics \cite{S4TechBook}.  
We mimic this effect in our simulation by placing a partially-scattering surface 
$50$~mm in front of the telescope's focal plane where the ray-bundles are sourced.  The scattering profile of this surface is modeled as a Lambertian, in which the intensity of the scattered radiation is given by $I = I_0 \cos{\theta}$, where $0 \leq \theta \leq \pi/2$ is the angle of the incident ray relative to the scattering element's surface normal and $I_0$ is a normalization factor. The simulations use a scattering fraction of 1\% based on the rough expectation of such wide-angle power leaving the receiver from the lab measurements \cite{S4TechBook} and the relative ease of scaling to other estimates.  We note that because the TMA mirrors are monolithic, it is not necessary to include any scattering or diffraction from gaps between mirror panels.  The mirror surfaces are assumed to be perfectly specularly reflecting; 
we ignore Ruze scattering due to the small-scale surface roughness of the mirrors (<1 microns \cite{Natoli}) because such scattering is much smaller than 1\% for three mirrors for $\lambda \ge 1$mm.

We use these simulations to explore the trade-offs between three different surface treatments on the enclosure walls, varying the relative contributions of absorption, reflection, and scattering. The three configurations are:
\begin{itemize}
    \item[(a)] a predominantly reflective enclosure, which specularly reflects 90\% of the incident power and absorbs the remaining 10\%. 
    \item[(b)] an enclosure with an ideal 100\% absorbing surface.
    \item[(c)] a predominantly scattering enclosure, for which 90\% of the incident rays scatter following a Lambertian profile and the remaining 10\% are absorbed. 
\end{itemize}
The 10\% absorption used in cases (a) and (c) allows the simulations to converge even when rays are trapped in trajectories that lead to an excessive number of bounces.

Highly reflective baffles have been used by a number of CMB instruments (e.g.\ \cite{Nagy_thesis, ACT_sidelobes2,SPT_George_etal}), but as we shall see below lead to significant sidelobe structure at wide angles for the design studied here. Highly absorptive baffles have also been used by some instruments \cite{Bicep3,CLASS,SPTpol_ref}; perfect absorption would eliminate the sidelobes we study here, but can cause undesirable optical loading on the detectors and reduced sensitivity if significant wide angle rays interact with that surface.  Such surfaces are also typically quite fragile, and the real absorption varies with incidence angle and photon frequency.  This motivates us to also study the scattering enclosure case to see if such a treatment can sufficiently reduce sidelobe structure, while maintaining low optical loading on the detectors.

\section{Results}
\label{sec: results}
\subsection{Sidelobe Maps}

\begin{figure}[htb]
 \begin{centering}
 \includegraphics[scale=0.315]{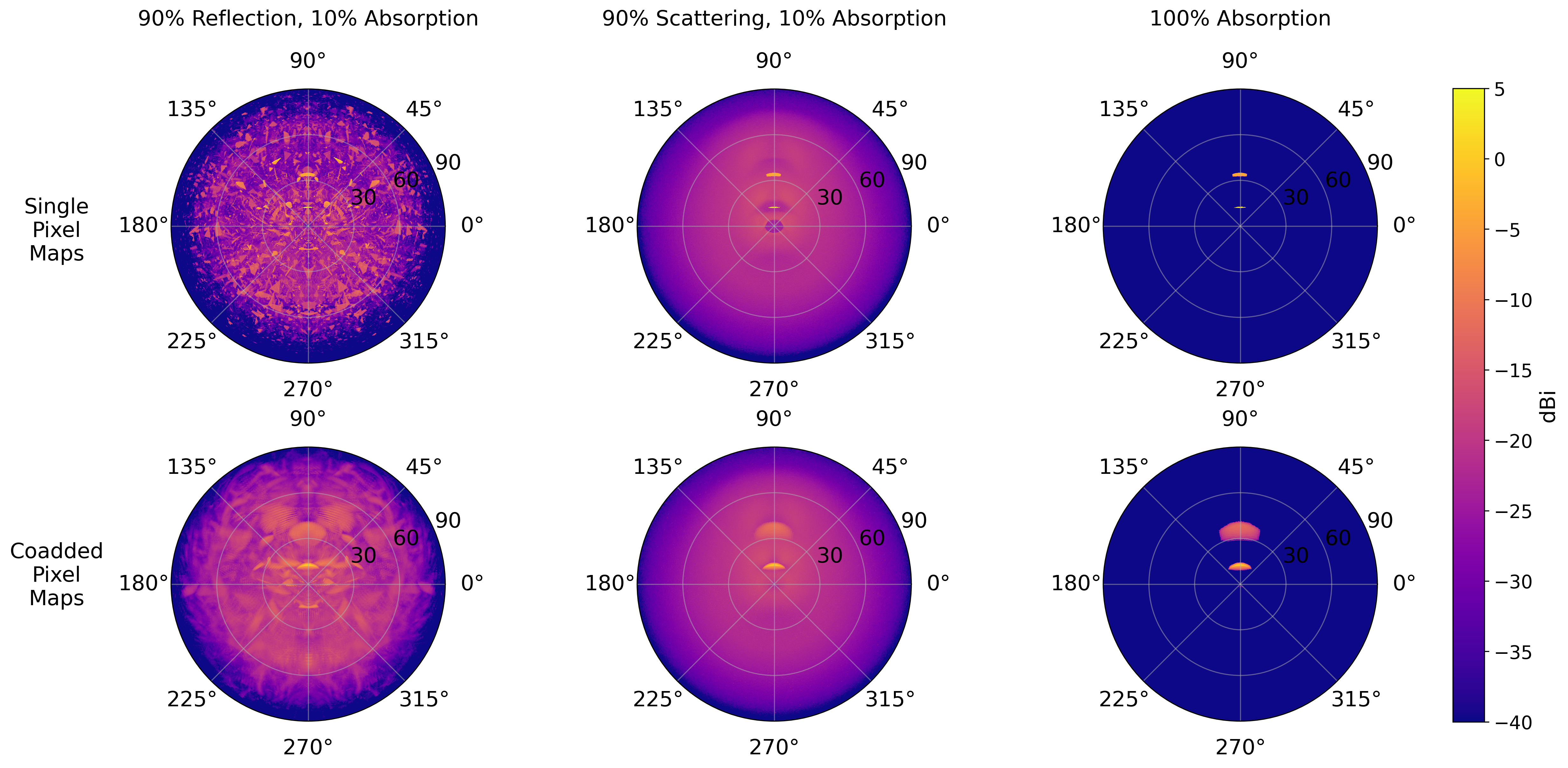}
 \par
 \end{centering}
 \caption{Beam maps produced by the ray tracing simulations showing the sidelobe power with different enclosure surface treatments.  The top row shows results for a single detector pixel at the center of the focal plane.  The bottom row shows coadded maps for 153 detector pixels distributed across the focal plane.  Specularly reflecting enclosure walls (left column) cause sharp sidelobe features.  Scattering enclosure walls (center column) blur those features, greatly lowering the sidelobe contrast at large angles.  Perfectly absorbing walls (right column) remove any features caused by scattered power interacting the enclosure walls, leaving only two features due to scattered power exiting the enclosure after bouncing off the mirrors.  While coadding blurs the sharp features from the reflective enclosure, the map is still less smooth than the scattering case.}
 \label{fig: TMA_maps}
 \end{figure}

For each of these three enclosure configurations, a beam map is made by launching a cone of $10^8$ rays from a specific location on the focal plane and recording the resulting far-field power distribution. Examples of such maps for a single detector pixel located at the center of the focal plane are shown in the top row of Figure \ref{fig: TMA_maps}.  A convergence test was performed increasing the number of rays by a factor of 10. Changes to the resulting power spectrum were less than one part in $10^5$ and are therefore negligible.

In the case of the absorbing enclosure, the geometry prevents the scattered rays from coupling directly to the sky, but reflection 
% from the primary and tertiary mirrors 
from the mirrors (scattered power to the tertiary, or directly to the primary) creates the two distinct bright patches.  The low power at wide angles comes at the cost of higher optical loading. Assuming a 300~K enclosure, the effective Rayleigh-Jeans temperature of the enclosure-wall-induced loading, for 1\% scattering at the receiver window, is 2.7~K.  This is reduced from the expected 3~K due to power which couples to the sky via the two bright sidelobes mentioned above.  
%the which is roughly a factor of 2 higher than for a reflective baffle. 
Since this adds undesirable loading on the detectors, and since real enclosure walls would not be perfectly absorbing at all incidence angles and observing frequencies (thereby inducing proportionally reduced versions of the specularly reflecting sidelobe pattern), we focus mainly on the sidelobe maps from specularly reflective and scattering enclosure walls.  
These both spread the power over a much wider range of angles, as shown in Figure \ref{fig: TMA_maps}, with the highly reflective surface creating much sharper features than the scattering surface.  
The output maps from Zemax are in units of Watts/sr;  we convert these to dBi by dividing by the Watts/sr for an isotropic radiator with the same total power as the sum of all the launched rays.

For any detector that is not centered on the focal plane, the main beam exits the telescope at an angle relative to the central pixel's beam.  CMB data analysis pipelines typically coadd maps made by pixels spread across the focal plane to produce a single sky map.  We replicate this effect in our simulations to find the effective sidelobes by similarly coadding 153 individual detector maps, rotated so the main beams are coincident.  The detector locations are distributed evenly on a 200~mm grid within a 2~m diameter circle, covering most of the approximately 2.5~m diameter diffraction-limited field of view at a wavelength of 1~mm.  The bottom row of Figure \ref{fig: TMA_maps} shows that coadding the pixels results in significant smoothing of the sharp, localized features seen in the single-pixel case.

\subsection{Ground Pick-up and Angular Power Spectra}
The maps presented in the previous section show that, for non-absorbing enclosure walls, significant sidelobe power at large angles could interact with the ground.  Our primary concern is the signal variations that would be caused by these sidelobes scanning across features in the unknown ``brightness map'' 
of the ground;  thus, it is not so much the level of the sidelobes, but rather the variations in the level of the sidelobes, that is of concern.

As our primary interest here is to avoid contamination of measurements of degree-scale features in the CMB, we are most interested in degree-scale 
contrast in the sidelobe maps.  A convenient measure of such contrast in a map $P(\theta,\phi)$ is 
found by calculating angular power spectra. Such power spectra are expressed as values of 
$C_\ell^{1/2} = (<a_{l,m} a_{l,m}>)^{1/2}$, where the average is over $m$, and the $a_{l,m}$'s are the coefficients of an expansion in spherical harmonics,  $P(\theta,\phi) = \sum_{l,m} a_{l,m} Y_{l,m}(\theta,\phi)$.  We note that a sidelobe pattern with minima and maxima separated by $1^\circ$ would appear at $\ell \sim 180$, with lower $\ell$ representing sidelobe variations at larger angular scales.  

\begin{figure}
 \begin{centering}
 \includegraphics[scale=0.65]{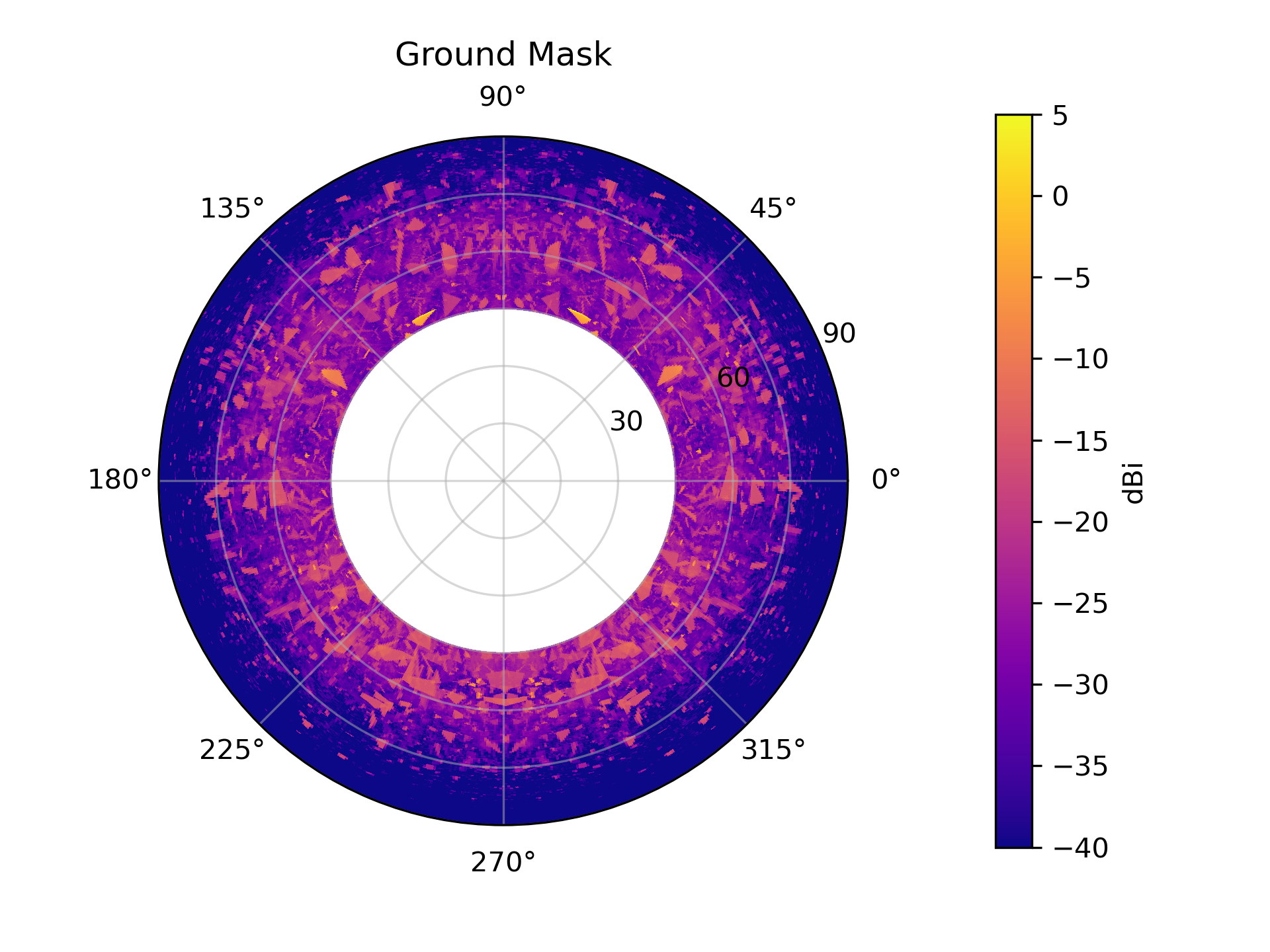}
 \par
 \end{centering}
 \caption{A masked sidelobe map showing the potential ground pick-up with the telescope at $45^\circ$ elevation and full boresight rotation.}
 \label{fig: Ground Mask}
 \end{figure}

A mask is applied to the maps to isolate the sidelobes that could be incident on the ground as shown in Figure \ref{fig: Ground Mask}. The mask is generated by assuming an observing elevation of $45^\circ$ and accounting for the possibility of $360^\circ$ of boresight rotation. We note that changing the observing elevation changes the radius of the inner (white) circle in the mask. 

Angular power spectra are then estimated from the masked maps using standard utilities from the \texttt{HEALPy} package \cite{Healpix}.  The maps are first rebinned into the \texttt{HEALPy} pixelization at nside 2048, and then the monopole is subtracted from each.  The power spectrum is produced by the \texttt{anafast} utility in \texttt{HEALPy} with the lmax parameter set to 400.  To check for artifacts from leakage between power spectrum bins, an input map was generated from a delta-function power spectrum at $\ell = 10$ and processed by the same pipeline.  Any signal observed in other bins of the processed power spectrum can be attributed to pipeline leakage, as described in \cite{transfer_matrix_paper}. In this case, the largest leakage signals were still at least a factor of 10 below than the lowest measured sidelobe power, so any leakage has a negligible impact on these results.

\begin{figure}
 \begin{centering}
 \includegraphics[scale=0.8]{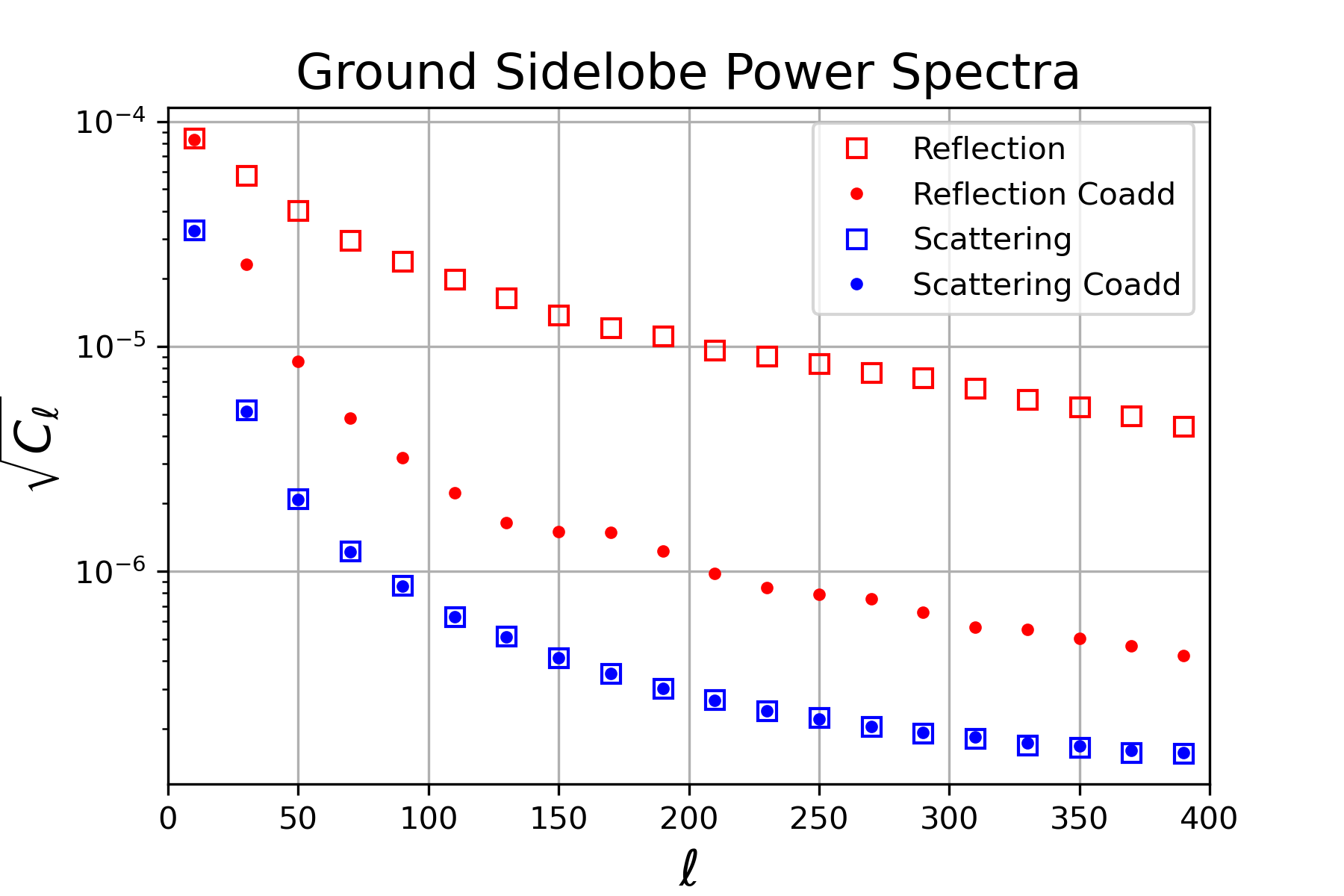}
 \par
 \end{centering}
 \caption{Angular power spectra ($C^{1/2}_\ell$ vs $\ell$) of single detector and coadded TMA sidelobe simulations at the degree angular scales ($50 < \ell < 300)$ relevant for measurements of primordial B-modes.  %Reflective enclosure walls (red) lead to sidelobes that are significantly worse than (in the single-detector case, red dots) or similar to (in the coadded focal plane case, filled red squares). 
 The highly scattering enclosure surface (blue open squares for the single detector case and blue dots for the coadded focal plane case) provide significantly lower sidelobe pick-up than the reflective surface cases.}
 \label{fig: power_spec}
 \end{figure}
 
The power spectra for the single detector and coadded maps are shown in Figure \ref{fig: power_spec}.  The blurring of the sharp features caused by coadding detectors across the focal plane results in lower ground pick-up on degree angular scales than the single pixel case.  This effect is largest for the reflective enclosure spectra, where the single pixel features are strong.  The scattering enclosure has significantly lower ground pick-up than the reflective case, for both single detector and coadded spectra. For this reason, the rest of this paper focuses on the scattering wall case.

An enclosure surface that is partially scattering and partially specularly reflecting will lead to a sidelobe angular power spectrum that is an appropriately weighted combination of the power spectra shown in Figure~\ref{fig: power_spec}.  To keep the residual impact of specular reflection below the scattering spectrum over the range $50 \lt \ell \lt 300$, the specular reflection needs to be less than $\sim 2$\% for a single detector, or $\sim 20$\% for a
fully coadded focal plane.  For a multi-frequency focal plane like that planned for CMB-S4 \cite{PatoSPIE}, with different frequency bands occupying different positions in the focal plane, the real criterion lies somewhere betweeen these two values. 
 For the rest of this paper we adopt the conservative target of keeping specular reflection less than 2\%.
 
\begin{figure}
 \begin{centering}
  \includegraphics[scale=0.45]{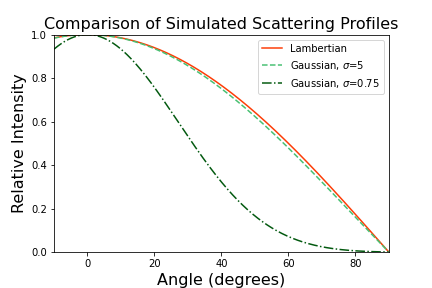}
 \includegraphics[scale=0.38]{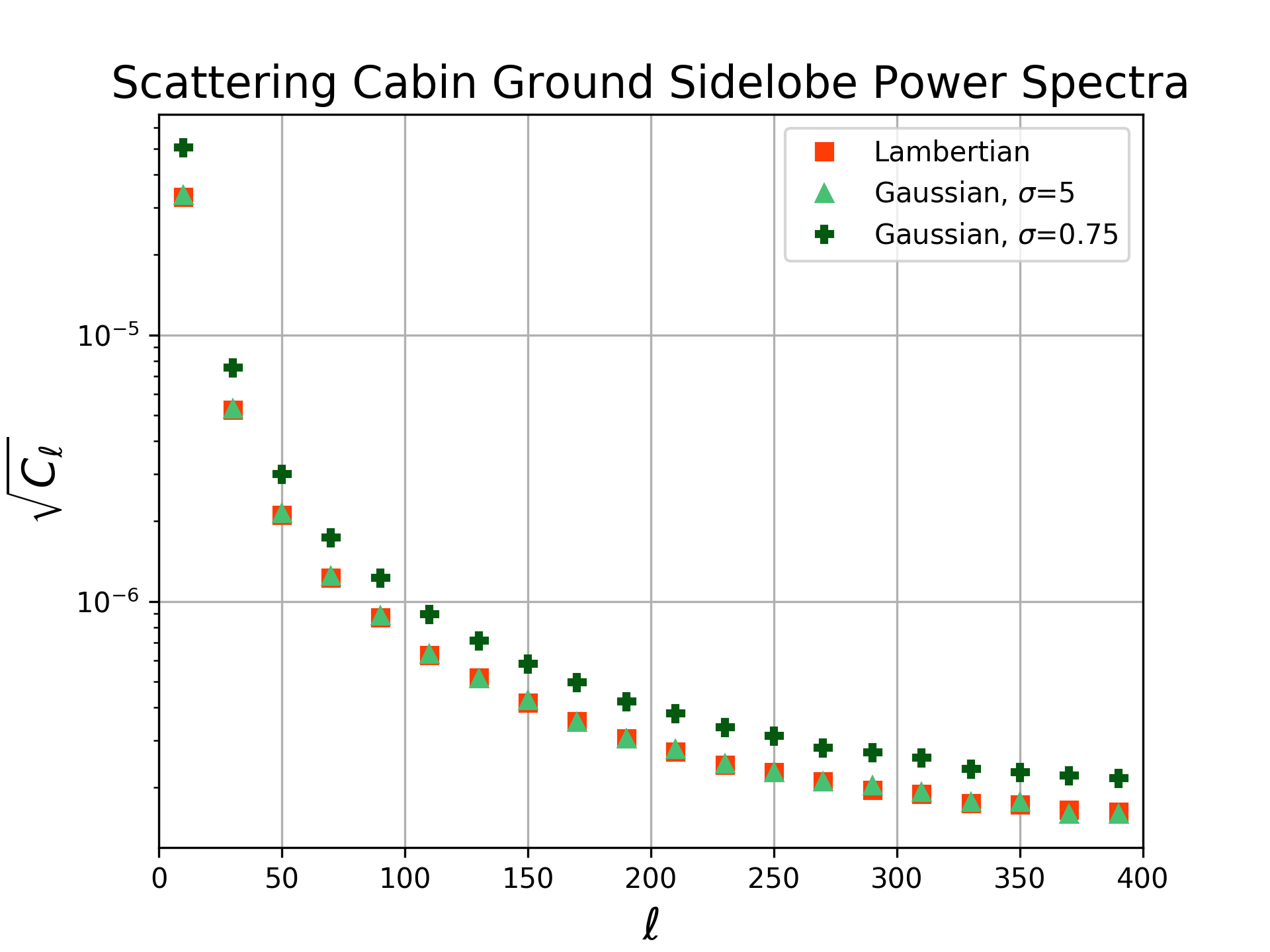}
 \par
 \end{centering}
 
 \caption{In the case of the highly scattering enclosure, the ground pick-up is relatively insensitive to the particular scattering profile assumed in the simulation.  The Gaussian profile characterized by $\sigma=0.75$ has a significantly narrower scattering profile than the Lambertian case, which is well approximated by $\sigma=5$ (left).  However, the expected ground pick-up from this profile is only a factor of 1.3 higher on degree angular scales (right). }
 \label{fig: Gauss_power_spectra}
 \end{figure}

 In our simulations above, we assumed a Lambertian scattering profile;   we tested the sensitivity of our results to the details of that profile using two Gaussian profiles illustrated in Figure \ref{fig: Gauss_power_spectra}. 
One of those profiles closely approximates the Lambertian, while the other is significantly narrower.  The resulting ground pick-up angular power spectrum for the narrower profile is only a factor of a few higher than the Lambertian case;  this is not surprising, given that most rays bounce multiple times before exiting the enclosure.  This relative insensitivity to the details of the scattering profile allows significant freedom in designing and implementing a suitable scattering enclosure wall surface.

\section{Prototyping a Scattering Surface}
\label{sec: proto_surfaces}
\subsection{Scattering Surface Design}
\label{sec: surf_design}
In addition to providing residual specular reflection below $\sim 2$\% as noted above, a few other considerations influence the design of a good candidate scattering wall material.  In particular, the scattering surface should work well over a very broad frequency range, in bands centered from 20 to 280~GHz.  Additionally, since the surface area to be covered is $\sim300$~m$^2$, we would like the material to be lightweight and easy to manufacture in large sheets.  The sheets must be mounted to the telescope structure in a way that is both mechanically robust and able to survive harsh ambient conditions at the South Pole for the full 15 year lifetime of the experiment.  Here we explore appropriately textured thin sheets of aluminum as a candidate scattering surface.

The scattering properties and gain loss of telescope mirrors that  have random Gaussian-shaped deformations with a given rms and coherence length are often modeled using the formulation developed by Ruze~\cite{Ruze_orig}.  One of the main findings of that work is that the antenna gain is reduced by  a factor $e^{-(4\pi\epsilon/\lambda)^2}$, where $\epsilon$ is the surface rms error and $\lambda$ is the wavelength.  The antenna gain here is equivalent to our fractional specular reflection;  this suggests that we need $e^{-(4\pi\epsilon/\lambda)^2} < 0.02$, which for our longest wavelength of interest (15~mm) leads to $\epsilon > 2.4$~mm.  Another finding is that the coherence length $c$ of the deformations sets the width of the scattered power pattern, with the width $\theta \propto \lambda/c$.  Given our desire for wide angle scattering, this suggests we need to keep $c \sim \lambda$, which would be daunting for operation near
$\lambda \sim 1$~mm for a surface with an rms of 2.4~mm.

However, we can explore surfaces with coherence lengths larger than $\lambda$ using ray tracing simulations, to see if they can provide satisfactory scattering to wide angles.  These simulations are also performed with the Zemax OpticsStudio software package in non-sequential mode. Nearly parallel ray bundles with incidence angles ranging from 0$^\circ$ to 2$^\circ$ are launched from a central point towards the scattering surface. The small variation of the incident ray angles compensates for the finite resolution of the imported CAD surfaces, which would otherwise generate an unphysical discretized scattering profile.  In these simulations, a 15~cm square scattering surface is modeled as 100\% reflective, and the scattering profile is found by averaging annularly around a hemispherical detector with 720 radial slices.  

After simulating a number of regular smooth feature patterns, we found that a surface constructed from spatially filtered Gaussian random ``noise'' provides excellent wide angle scattering performance.  A numerical model of the simulated surface was constructed by generating Gaussian random heights 
%with rms = 0.58~mm 
on a 15~cm square grid with a pixel size of 0.5~mm, and spatially filtered in Fourier space to remove modes with spatial wavelengths less than 10 mm. After filtering, the surface heights %are multiplied by a scale factor of approximately 3.38 
were scaled to achieve the final filtered surface rms of 1.78~mm for the prototype design described in this paper.  This is slightly lower than the 2.4~mm rms suggested above, but satisfactory for the $\lambda \sim 3$~mm measurements described below. 

\subsection{Prototype Fabrication and Characterization}
\begin{figure}
 \begin{centering}
 \includegraphics[scale=0.08, angle =90]{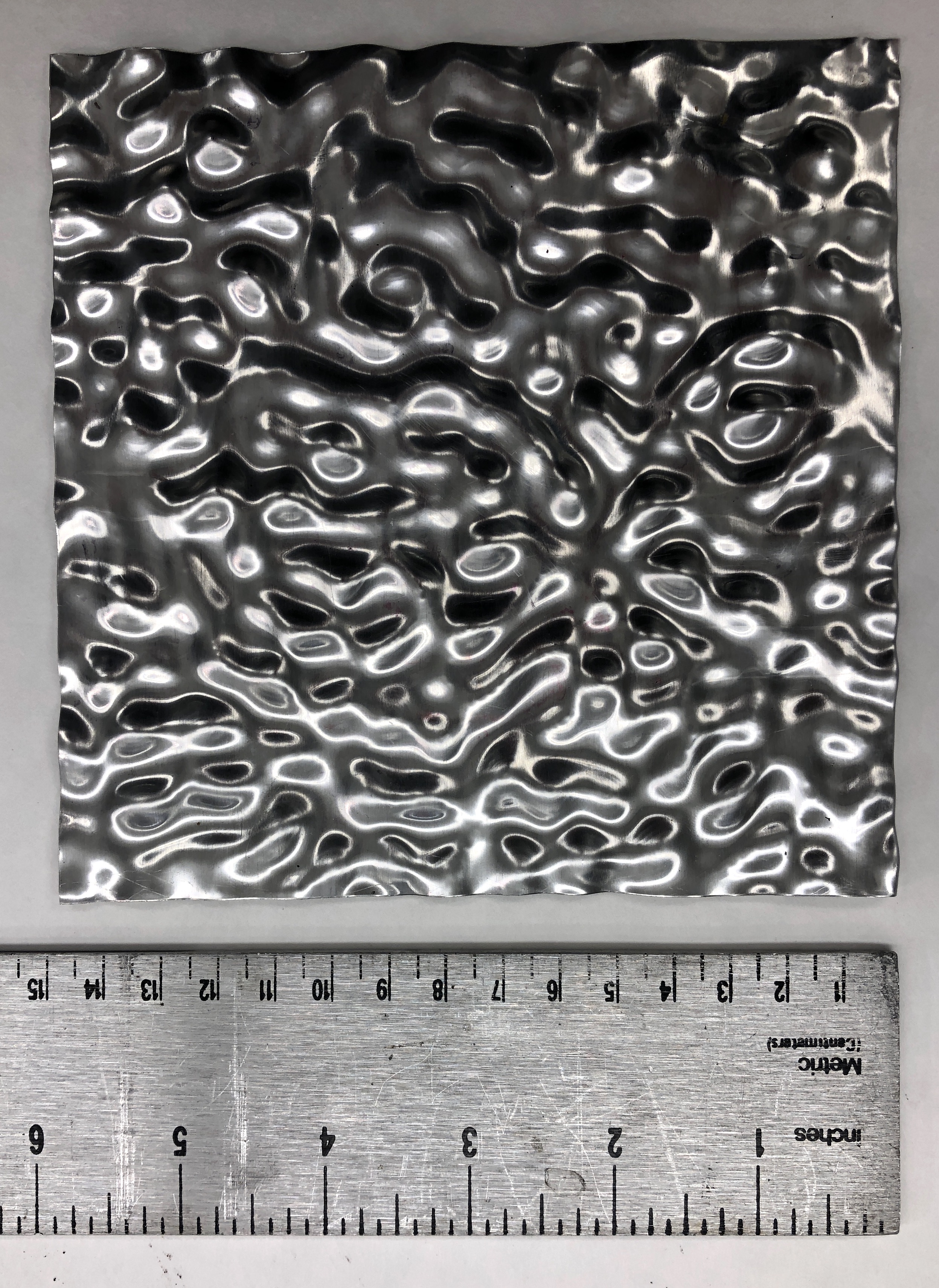}
 \par
 \end{centering}
 \caption{A photograph of the prototype aluminum scattering surface.  }
 \label{fig: prototype}
 \end{figure}

A prototype random noise surface was fabricated from a 150 mm square of pressed aluminum as shown in Figure \ref{fig: prototype}. The surface features were defined by a negative mold 3D printed from polylactic acid (PLA).  A 0.5~mm thick aluminum 1100 sheet was then pressed onto the mold using a hydraulic press with approximately 1000~psi. Even pressure distribution was facilitated by placing rubber sheets between the piston side of the press and the aluminum sheet, and a spray-on wax was used to lubricate the aluminum surface to reduce shearing forces. 3D laser scanner measurements allowed the protoype surface rms to be compared to the design; the measured features have an rms of 1.81~mm, which is in good agreement with the 1.78~mm model rms.

%\subsection{Comparing the Model and Prototype}
%including spinning measurements
\begin{figure}
 \begin{centering}
 \includegraphics[scale=0.5]{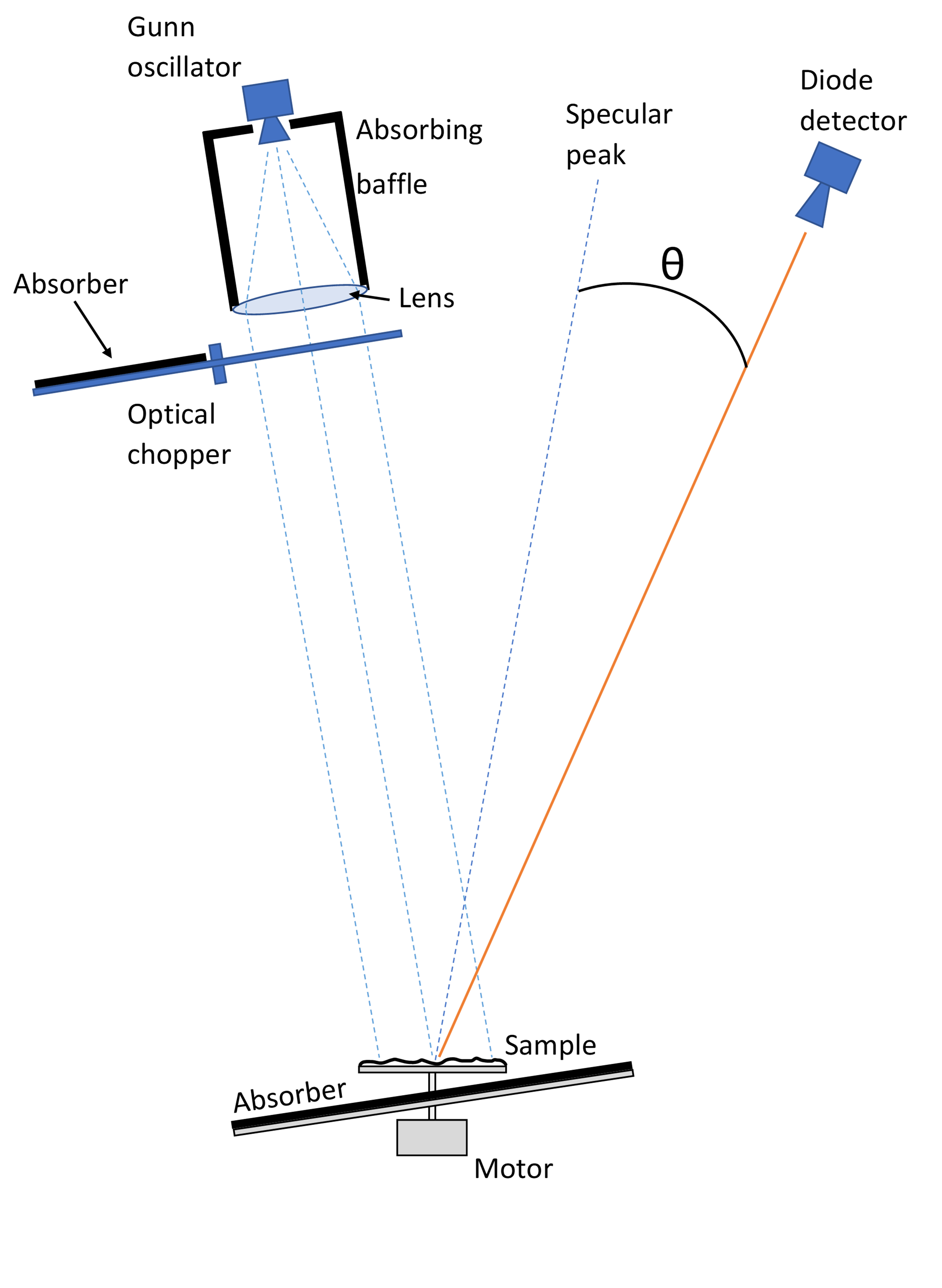}
 \par
 \end{centering}
 \caption{A diagram of the apparatus used to measure the prototype scattering profile.}
 \label{fig: Gunn_diagram}
 \end{figure}

The prototype's scattering profile was measured with the apparatus shown in Figure \ref{fig: Gunn_diagram}. A Gunn oscillator generates a signal at a discrete, tunable frequency ranging from 77-112~GHz.  The signal is broadcast by a horn at the focus of a plastic lens, creating a 10~cm diameter parallel beam.  That beam passes through an absorbing-blade optical chopper before interacting with the sample.  The sample was trimmed to a 150~mm diameter circle and mounted on an aluminum plate surrounded by absorbing material. The sample is tilted at an angle of roughly $ 3^\circ$ relative to the incoming beam to allow detector access to the specular reflection peak.  Power leaving the sample is measured by a diode detector mounted on a swing-arm to keep it pointed directly at the sample at all angles.  The diode signal is sent to a lock-in amplifier for synchronous detection, with the reference taken from the optical chopper. 

\begin{figure}
 \begin{centering}
 \includegraphics[scale=0.4]{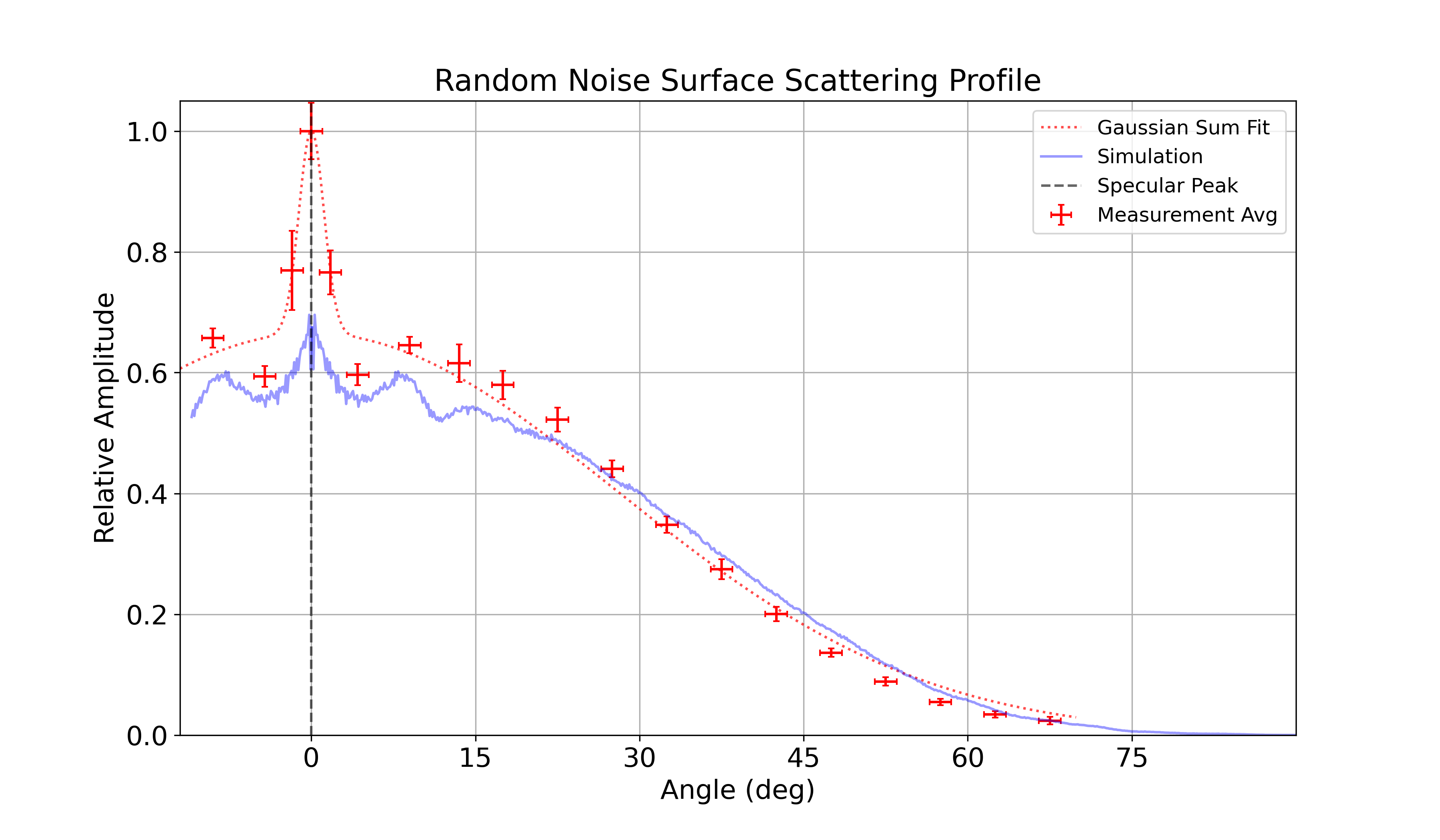}
 \par
 \end{centering}
 \caption{The measured (red data points) scattering profile of the prototype random noise surface compared to the prediction from ray tracing simulations (blue line). The measurement is the average of 10 frequencies from 77 GHz - 112 GHz. The widths of the simulated and measured distributions are similar. 
 The measurement shows a significantly higher amplitude specular peak than the simulation.  The  red dotted line shows a fit to the measured points using a sum of two Gaussians, one of which is broad (the scattered power, FWHM=69$^\circ$), and one narrow (the specular reflection, FWHM=2.7$^\circ$).  Integrating each of these two-dimensional Gaussians shows that the specular contains only $\sim 0.1$\% of the total power, which is well below the maximum desired $\sim 2 \%$.}
 \label{fig: noise_surface_meas}
 \end{figure}

To minimize speckle effects due to the long coherence length of the Gunn oscillator, we spun the sample on its axis, rapidly relative to the lock-in time constant of 1 second.  This provides an annular average over the scattering pattern and speckle effects, reducing the noise in the measured scattering profile shown as the red data points in 
Figure \ref{fig: noise_surface_meas}. We note that in broadband CMB measurements the coherence length is much smaller, and the inherent spectral averaging naturally mitigates such effects. The data points and errorbars are constructed from scans at 10 discrete frequencies between 77~GHz and 112~GHz, where we normalize each scan relative to the others by the total integrated power before averaging.  For comparison, the Zemax raytracing simulations described in Section \ref{sec: surf_design} are shown as the blue line.  Both the width and shape agree reasonably well with the measurement. Finally, we fit the data points to a sum of two two-dimensional Gaussians, one broad to represent the scattered power, and one narrow to represent the specularly reflected peak.  The resulting fit is shown as the red dotted line;  integrating each two-dimensional Gaussian separately indicates that roughly 0.1\% of the power lies in the specular peak, well below the 2\% maximum desired. We attribute the specular peak in the measurement to flat areas in the aluminum which didn't perfectly conform to the 3D printed mold. Based on these simulations and measurements, this ``noise surface'' meets our scattering criteria near 90~GHz.

%Looking at the surface derivatives reveals that 3\% of the manufactured surface is $\geq45^\circ$ whereas 9\% of the model surface is $\geq45^\circ$.

\subsection{Enclosure Mounting}
The structure of the telescope enclosure will be formed by carbon fiber hexagonal honeycomb or foam-core panels supported by a carbon fiber truss.  Rigidly mounting the aluminum scattering panels is thus challenging due to the expected difference in the thermal contraction of aluminum relative to carbon fiber over the range of temperatures expected at the South Pole, given the relatively large size of the panels needed for efficient installation.  The mechanical mounting scheme also needs to remain robust after many cycles of wind deformation experienced over the telescope lifetime.

A series of thermal and mechanical tests were performed to identify a method of attaching the aluminum scattering panels that satisfies all these conditions.  In one method, Loctite E-120HP epoxy was used to bond stainless steel fasteners to the carbon fiber surface, chosen because it has been successfully used by other experiments to bond carbon fiber to aluminum (e.g.\ \cite{XLCal}).  Although a series of simulated wind loading tests demonstrated that the bonds could survive tens of thousands of deformation cycles at temperatures ranging from 20 to -70 \degree C, they did not survive the stresses of differential thermal contraction over the same range over length scales of 1.2~m.  To avoid these bonds altogether, oversized holes were drilled through both the carbon fiber and aluminum panels and secured with stainless steel fasteners.  The panels remained robustly attached after multiple thermal cycles to -70 \degree C, and the simplicity of this mounting method would allow for easy assembly at the South Pole site.  

\section{Conclusions}
Controlling degree-scale variations in far sidelobes that would generate ground pick-up is critical for experiments searching for signals from inflation in CMB B-mode polarization.  In large aperture telescopes, such sidelobe structure could be generated by power that spills over the mirrors and interacts with the surrounding enclosure walls.  We have shown here that the degree-scale variations in such sidelobes can be greatly reduced by scattering rather than specularly reflecting that power.  Our simulations suggest the ground pick-up with a scattering enclosure would be significantly lower than has been observed by current generation telescopes, representing a path to reducing the systematic error contribution in future, more sensitive instruments. We have also presented a practical solution for fabricating small prototypes that meet the general requirements for such scattering panels.  Future work will explore fabrication methods to produce larger such panels at reasonable cost and testing the scattering performance at a broader range of frequencies.

\section*{Acknowledgements}
JER acknowledges support from DOE HEP award DE-SC0021434 and NSF award 1935892. Work at WUSTL was supported by DOE HEP award DE-SC0017987.

\section*{Disclosures}
The authors declare no conflicts of interest.

\section*{Data availability} Data underlying the results presented in this paper are available in Ref. \cite{PublicDataRepository}.

%%%%%%%%%% If using BibTeX:
\bibliography{references}

\begin{thebibliography}{10}
\newcommand{\enquote}[1]{``#1''}

\bibitem{BK_r}
P.~A.~R. {Ade}, Z.~{Ahmed}, M.~{Amiri}, D.~{Barkats}, R.~B. {Thakur}, C.~A.
  {Bischoff}, D.~{Beck}, J.~J. {Bock}, H.~{Boenish}, E.~{Bullock}, V.~{Buza},
  J.~R. {Cheshire}, J.~{Connors}, J.~{Cornelison}, M.~{Crumrine},
  A.~{Cukierman}, E.~V. {Denison}, M.~{Dierickx}, L.~{Duband}, M.~{Eiben},
  S.~{Fatigoni}, J.~P. {Filippini}, S.~{Fliescher}, N.~{Goeckner-Wald}, D.~C.
  {Goldfinger}, J.~{Grayson}, P.~{Grimes}, G.~{Hall}, G.~{Halal}, M.~{Halpern},
  E.~{Hand}, S.~{Harrison}, S.~{Henderson}, S.~R. {Hildebrandt}, G.~C.
  {Hilton}, J.~{Hubmayr}, H.~{Hui}, K.~D. {Irwin}, J.~{Kang}, K.~S. {Karkare},
  E.~{Karpel}, S.~{Kefeli}, S.~A. {Kernasovskiy}, J.~M. {Kovac}, C.~L. {Kuo},
  K.~{Lau}, E.~M. {Leitch}, A.~{Lennox}, K.~G. {Megerian}, L.~{Minutolo},
  L.~{Moncelsi}, Y.~{Nakato}, T.~{Namikawa}, H.~T. {Nguyen}, R.~{O'Brient},
  R.~W. {Ogburn}, S.~{Palladino}, T.~{Prouve}, C.~{Pryke}, B.~{Racine}, C.~D.
  {Reintsema}, S.~{Richter}, A.~{Schillaci}, R.~{Schwarz}, B.~L. {Schmitt},
  C.~D. {Sheehy}, A.~{Soliman}, T.~S. {Germaine}, B.~{Steinbach}, R.~V.
  {Sudiwala}, G.~P. {Teply}, K.~L. {Thompson}, J.~E. {Tolan}, C.~{Tucker},
  A.~D. {Turner}, C.~{Umilt{\`a}}, C.~{Verg{\`e}s}, A.~G. {Vieregg},
  A.~{Wandui}, A.~C. {Weber}, D.~V. {Wiebe}, J.~{Willmert}, C.~L. {Wong},
  W.~L.~K. {Wu}, H.~{Yang}, K.~W. {Yoon}, E.~{Young}, C.~{Yu}, L.~{Zeng},
  C.~{Zhang}, S.~{Zhang}, and {Bicep/Keck Collaboration}, \enquote{{Improved
  Constraints on Primordial Gravitational Waves using Planck, WMAP, and
  BICEP/Keck Observations through the 2018 Observing Season},}
  {\protect\JournalTitle{Physical Review Letters}} \textbf{127}, 151301 (2021).

\bibitem{Spider_r}
{SPIDER Collaboration}, P.~A.~R. {Ade}, M.~{Amiri}, S.~J. {Benton}, A.~S.
  {Bergman}, R.~{Bihary}, J.~J. {Bock}, J.~R. {Bond}, J.~A. {Bonetti}, S.~A.
  {Bryan}, H.~C. {Chiang}, C.~R. {Contaldi}, O.~{Dor{\'e}}, A.~J.
  {Duivenvoorden}, H.~K. {Eriksen}, M.~{Farhang}, J.~P. {Filippini}, A.~A.
  {Fraisse}, K.~{Freese}, M.~{Galloway}, A.~E. {Gambrel}, N.~N. {Gandilo},
  K.~{Ganga}, R.~{Gualtieri}, J.~E. {Gudmundsson}, M.~{Halpern}, J.~{Hartley},
  M.~{Hasselfield}, G.~{Hilton}, W.~{Holmes}, V.~V. {Hristov}, Z.~{Huang},
  K.~D. {Irwin}, W.~C. {Jones}, A.~{Karakci}, C.~L. {Kuo}, Z.~D. {Kermish},
  J.~S.~Y. {Leung}, S.~{Li}, D.~S.~Y. {Mak}, P.~V. {Mason}, K.~{Megerian},
  L.~{Moncelsi}, T.~A. {Morford}, J.~M. {Nagy}, C.~B. {Netterfield},
  M.~{Nolta}, R.~{O'Brient}, B.~{Osherson}, I.~L. {Padilla}, B.~{Racine}, A.~S.
  {Rahlin}, C.~{Reintsema}, J.~E. {Ruhl}, M.~C. {Runyan}, T.~M. {Ruud}, J.~A.
  {Shariff}, E.~C. {Shaw}, C.~{Shiu}, J.~D. {Soler}, X.~{Song}, A.~{Trangsrud},
  C.~{Tucker}, R.~S. {Tucker}, A.~D. {Turner}, J.~F. {van der List}, A.~C.
  {Weber}, I.~K. {Wehus}, S.~{Wen}, D.~V. {Wiebe}, and E.~Y. {Young},
  \enquote{{A Constraint on Primordial $B$-Modes from the First Flight of the
  SPIDER Balloon-Borne Telescope},} {\protect\JournalTitle{The Astrophysical
  Journal}} \textbf{927}, 174 (2022).

\bibitem{Planck_r}
M.~{Tristram}, A.~J. {Banday}, K.~M. {G{\'o}rski}, R.~{Keskitalo}, C.~R.
  {Lawrence}, K.~J. {Andersen}, R.~B. {Barreiro}, J.~{Borrill}, H.~K.
  {Eriksen}, R.~{Fernandez-Cobos}, T.~S. {Kisner},
  E.~{Mart{\'\i}nez-Gonz{\'a}lez}, B.~{Partridge}, D.~{Scott}, T.~L.
  {Svalheim}, H.~{Thommesen}, and I.~K. {Wehus}, \enquote{{Planck constraints
  on the tensor-to-scalar ratio},} {\protect\JournalTitle{Astronomy \&
  Astrophysics}} \textbf{647}, A128 (2021).

\bibitem{SPT_r}
J.~T. {Sayre}, C.~L. {Reichardt}, J.~W. {Henning}, P.~A.~R. {Ade}, A.~J.
  {Anderson}, J.~E. {Austermann}, J.~S. {Avva}, J.~A. {Beall}, A.~N. {Bender},
  B.~A. {Benson}, F.~{Bianchini}, L.~E. {Bleem}, J.~E. {Carlstrom}, C.~L.
  {Chang}, P.~{Chaubal}, H.~C. {Chiang}, R.~{Citron}, C.~{Corbett Moran}, T.~M.
  {Crawford}, A.~T. {Crites}, T.~{de Haan}, M.~A. {Dobbs}, W.~{Everett},
  J.~{Gallicchio}, E.~M. {George}, A.~{Gilbert}, N.~{Gupta}, N.~W. {Halverson},
  N.~{Harrington}, G.~C. {Hilton}, G.~P. {Holder}, W.~L. {Holzapfel}, J.~D.
  {Hrubes}, N.~{Huang}, J.~{Hubmayr}, K.~D. {Irwin}, L.~{Knox}, A.~T. {Lee},
  D.~{Li}, A.~{Lowitz}, J.~J. {McMahon}, S.~S. {Meyer}, L.~M. {Mocanu},
  J.~{Montgomery}, A.~{Nadolski}, T.~{Natoli}, J.~P. {Nibarger}, G.~{Noble},
  V.~{Novosad}, S.~{Padin}, S.~{Patil}, C.~{Pryke}, J.~E. {Ruhl}, B.~R.
  {Saliwanchik}, K.~K. {Schaffer}, C.~{Sievers}, G.~{Smecher}, A.~A. {Stark},
  C.~{Tucker}, K.~{Vanderlinde}, T.~{Veach}, J.~D. {Vieira}, G.~{Wang},
  N.~{Whitehorn}, W.~L.~K. {Wu}, V.~{Yefremenko}, and {SPTpol Collaboration},
  \enquote{{Measurements of B-mode polarization of the cosmic microwave
  background from 500 square degrees of SPTpol data},}
  {\protect\JournalTitle{Physical Review D}} \textbf{101}, 122003 (2020).

\bibitem{S4_ref_design}
K.~Abazajian, G.~Addison, P.~Adshead, Z.~Ahmed, S.~W. Allen, D.~Alonso,
  M.~Alvarez, A.~Anderson, K.~S. Arnold, C.~Baccigalupi \emph{et~al.},
  \enquote{Cmb-s4 science case, reference design, and project plan,}
  {\protect\JournalTitle{arXiv preprint arXiv:1907.04473}}  (2019).

\bibitem{S4_r_forecast}
K.~{Abazajian}, G.~E. {Addison}, P.~{Adshead}, Z.~{Ahmed}, D.~{Akerib},
  A.~{Ali}, S.~W. {Allen}, D.~{Alonso}, M.~{Alvarez}, M.~A. {Amin},
  A.~{Anderson}, K.~S. {Arnold}, P.~{Ashton}, C.~{Baccigalupi}, D.~{Bard},
  D.~{Barkats}, D.~{Barron}, P.~S. {Barry}, J.~G. {Bartlett}, R.~{Basu Thakur},
  N.~{Battaglia}, R.~{Bean}, C.~{Bebek}, A.~N. {Bender}, B.~A. {Benson},
  F.~{Bianchini}, C.~A. {Bischoff}, L.~{Bleem}, J.~J. {Bock}, S.~{Bocquet},
  K.~K. {Boddy}, J.~{Richard Bond}, J.~{Borrill}, F.~R. {Bouchet},
  T.~{Brinckmann}, M.~L. {Brown}, S.~{Bryan}, V.~{Buza}, K.~{Byrum},
  C.~{Hervias Caimapo}, E.~{Calabrese}, V.~{Calafut}, R.~{Caldwell}, J.~E.
  {Carlstrom}, J.~{Carron}, T.~{Cecil}, A.~{Challinor}, C.~L. {Chang},
  Y.~{Chinone}, H.-M. {Sherry Cho}, A.~{Cooray}, W.~{Coulton}, T.~M.
  {Crawford}, A.~{Crites}, A.~{Cukierman}, F.-Y. {Cyr-Racine}, T.~{de Haan},
  J.~{Delabrouille}, M.~{Devlin}, E.~{Di Valentino}, M.~{Dierickx}, M.~{Dobbs},
  S.~{Duff}, C.~{Dvorkin}, J.~{Eimer}, T.~{Elleflot}, J.~{Errard},
  T.~{Essinger-Hileman}, G.~{Fabbian}, C.~{Feng}, S.~{Ferraro}, J.~P.
  {Filippini}, R.~{Flauger}, B.~{Flaugher}, A.~A. {Fraisse}, A.~{Frolov},
  N.~{Galitzki}, P.~A. {Gallardo}, S.~{Galli}, K.~{Ganga}, M.~{Gerbino},
  V.~{Gluscevic}, N.~{Goeckner-Wald}, D.~{Green}, D.~{Grin}, E.~{Grohs},
  R.~{Gualtieri}, J.~E. {Gudmundsson}, I.~{Gullett}, N.~{Gupta}, S.~{Habib},
  M.~{Halpern}, N.~W. {Halverson}, S.~{Hanany}, K.~{Harrington}, M.~{Hasegawa},
  M.~{Hasselfield}, M.~{Hazumi}, K.~{Heitmann}, S.~{Henderson}, B.~{Hensley},
  C.~{Hill}, J.~{Colin Hill}, R.~{Hlo{\v{z}}ek}, S.-P. {Patty Ho}, T.~{Hoang},
  G.~{Holder}, W.~{Holzapfel}, J.~{Hood}, J.~{Hubmayr}, K.~M. {Huffenberger},
  H.~{Hui}, K.~{Irwin}, O.~{Jeong}, B.~R. {Johnson}, W.~C. {Jones}, J.~{Hwan
  Kang}, K.~S. {Karkare}, N.~{Katayama}, R.~{Keskitalo}, T.~{Kisner},
  L.~{Knox}, B.~J. {Koopman}, A.~{Kosowsky}, J.~{Kovac}, E.~D. {Kovetz},
  S.~{Kuhlmann}, C.-l. {Kuo}, A.~{Kusaka}, A.~{L{\"a}hteenm{\"a}ki}, C.~R.
  {Lawrence}, A.~T. {Lee}, A.~{Lewis}, D.~{Li}, E.~{Linder}, M.~{Loverde},
  A.~{Lowitz}, P.~{Lubin}, M.~S. {Madhavacheril}, A.~{Mantz}, G.~{Marques},
  F.~{Matsuda}, P.~{Mauskopf}, H.~{McCarrick}, J.~{McMahon}, P.~{Daniel
  Meerburg}, J.-B. {Melin}, F.~{Menanteau}, J.~{Meyers}, M.~{Millea},
  J.~{Mohr}, L.~{Moncelsi}, M.~{Monzani}, T.~{Mroczkowski}, S.~{Mukherjee},
  J.~{Nagy}, T.~{Namikawa}, F.~{Nati}, T.~{Natoli}, L.~{Newburgh}, M.~D.
  {Niemack}, H.~{Nishino}, B.~{Nord}, V.~{Novosad}, R.~{O'Brient}, S.~{Padin},
  S.~{Palladino}, B.~{Partridge}, D.~{Petravick}, E.~{Pierpaoli},
  L.~{Pogosian}, K.~{Prabhu}, C.~{Pryke}, G.~{Puglisi}, B.~{Racine},
  A.~{Rahlin}, M.~{Sathyanarayana Rao}, M.~{Raveri}, C.~L. {Reichardt},
  M.~{Remazeilles}, G.~{Rocha}, N.~A. {Roe}, A.~{Roy}, J.~E. {Ruhl},
  M.~{Salatino}, B.~{Saliwanchik}, E.~{Schaan}, A.~{Schillaci}, B.~{Schmitt},
  M.~M. {Schmittfull}, D.~{Scott}, N.~{Sehgal}, S.~{Shandera}, B.~D. {Sherwin},
  E.~{Shirokoff}, S.~M. {Simon}, A.~{Slosar}, D.~{Spergel}, T.~{St. Germaine},
  S.~T. {Staggs}, A.~{Stark}, G.~D. {Starkman}, R.~{Stompor}, C.~{Stoughton},
  A.~{Suzuki}, O.~{Tajima}, G.~P. {Teply}, K.~{Thompson}, B.~{Thorne},
  P.~{Timbie}, M.~{Tomasi}, M.~{Tristram}, G.~{Tucker}, C.~{Umilt{\`a}},
  A.~{van Engelen}, E.~M. {Vavagiakis}, J.~D. {Vieira}, A.~G. {Vieregg},
  K.~{Wagoner}, B.~{Wallisch}, G.~{Wang}, S.~{Watson}, B.~{Westbrook},
  N.~{Whitehorn}, E.~J. {Wollack}, W.~L. {Kimmy Wu}, Z.~{Xu}, H.~Y. {Eric
  Yang}, S.~{Yasini}, V.~G. {Yefremenko}, K.~{Won Yoon}, E.~{Young}, C.~{Yu},
  A.~{Zonca}, and {CMB-S4 Collaboration}, \enquote{{CMB-S4: Forecasting
  Constraints on Primordial Gravitational Waves},} {\protect\JournalTitle{The
  Astrophysical Journal}} \textbf{926}, 54 (2022).

\bibitem{padinTMA}
S.~Padin, \enquote{Three-mirror anastigmat for cosmic microwave background
  observations,} {\protect\JournalTitle{Applied optics}} \textbf{57},
  2314--2326 (2018).

\bibitem{PatoSPIE}
P.~A. {Gallardo}, B.~{Benson}, J.~{Carlstrom}, S.~R. {Dicker}, N.~{Emerson},
  J.~E. {Gudmundsson}, R.~{Hills}, M.~{Limon}, J.~{McMahon}, M.~D. {Niemack},
  J.~M. {Nagy}, S.~{Padin}, J.~{Ruhl}, and S.~M. {Simon}, \enquote{{Optical
  design concept of the CMB-S4 large-aperture telescopes and cameras},} in
  \emph{Millimeter, Submillimeter, and Far-Infrared Detectors and
  Instrumentation for Astronomy XI,}  vol. 12190 of \emph{Society of
  Photo-Optical Instrumentation Engineers (SPIE) Conference Series}
  J.~{Zmuidzinas} and J.-R. {Gao}, eds. (2022), p. 121900C.

\bibitem{SPT_George_etal}
E.~M. George, P.~Ade, K.~A. Aird, J.~E. Austermann, J.~A. Beall, D.~Becker,
  A.~Bender, B.~A. Benson, L.~E. Bleem, J.~Britton, J.~E. Carlstrom, C.~L.
  Chang, H.~C. Chiang, H.-M. Cho, T.~M. Crawford, A.~T. Crites, A.~Datesman,
  T.~de~Haan, M.~A. Dobbs, W.~Everett, A.~Ewall-Wice, N.~W. Halverson,
  N.~Harrington, J.~W. Henning, G.~C. Hilton, W.~L. Holzapfel, S.~Hoover,
  N.~Huang, J.~Hubmayr, K.~D. Irwin, M.~Karfunkle, R.~Keisler, J.~Kennedy,
  A.~T. Lee, E.~Leitch, D.~Li, M.~Lueker, D.~P. Marrone, J.~J. McMahon,
  J.~Mehl, S.~S. Meyer, J.~Montgomery, T.~E. Montroy, J.~Nagy, T.~Natoli, J.~P.
  Nibarger, M.~D. Niemack, V.~Novosad, S.~Padin, C.~Pryke, C.~L. Reichardt,
  J.~E. Ruhl, B.~R. Saliwanchik, J.~T. Sayre, K.~K. Schaffer, E.~Shirokoff,
  K.~Story, C.~Tucker, K.~Vanderlinde, J.~D. Vieira, G.~Wang, R.~Williamson,
  V.~Yefremenko, K.~W. Yoon, and E.~Young, \enquote{Performance and on-sky
  optical characterization of the {SPTpol} instrument,} in \emph{{SPIE}
  Proceedings,}  W.~S. Holland, ed. ({SPIE}, 2012).

\bibitem{ACT_sidelobes}
P.~Gallardo, R.~D{\"u}nner, E.~Wollack, F.~Henriquez, and C.~Jerez-Hanckes,
  \enquote{Mirror illumination and spillover measurements of the atacama
  cosmology telescope,} in \emph{Millimeter, Submillimeter, and Far-Infrared
  Detectors and Instrumentation for Astronomy VI,}  vol. 8452 (International
  Society for Optics and Photonics, 2012), p. 845224.

\bibitem{ACT_sidelobes2}
P.~A. Gallardo, N.~F. Cothard, R.~Puddu, R.~D{\"u}nner, B.~J. Koopman, M.~D.
  Niemack, S.~M. Simon, and E.~J. Wollack, \enquote{Far sidelobes from baffles
  and telescope support structures in the atacama cosmology telescope,} in
  \emph{Millimeter, Submillimeter, and Far-Infrared Detectors and
  Instrumentation for Astronomy IX,}  vol. 10708 (International Society for
  Optics and Photonics, 2018), p. 107082L.

\bibitem{SO_optics_modeling}
J.~E. Gudmundsson, P.~A. Gallardo, R.~Puddu, S.~R. Dicker, A.~E. Adler, A.~M.
  Ali, A.~Bazarko, G.~E. Chesmore, G.~Coppi, N.~F. Cothard \emph{et~al.},
  \enquote{The simons observatory: modeling optical systematics in the large
  aperture telescope,} {\protect\JournalTitle{Applied Optics}} \textbf{60},
  823--837 (2021).

\bibitem{Adler_SPIE}
A.~E. {Adler} and J.~E. {Gudmundsson}, \enquote{{Modeling sidelobe response for
  ground-based mm-wavelength telescopes with the geometrical theory of
  diffraction},} in \emph{Society of Photo-Optical Instrumentation Engineers
  (SPIE) Conference Series,}  vol. 11453 of \emph{Society of Photo-Optical
  Instrumentation Engineers (SPIE) Conference Series} (2020), p. 114534O.

\bibitem{Wang:21}
C.-L. Wang, C.-P. Chiu, P.-J. Huang, S.-Y. Wang, and M.-J. Wang,
  \enquote{High-performance 1--10 thz integrating sphere,}
  {\protect\JournalTitle{Appl. Opt.}} \textbf{60}, 3784--3790 (2021).

\bibitem{Shirley:88}
L.~G. Shirley and N.~George, \enquote{Diffuser radiation patterns over a large
  dynamic range. 1: Strong diffusers,} {\protect\JournalTitle{Appl. Opt.}}
  \textbf{27}, 1850--1861 (1988).

\bibitem{S4TechBook}
M.~H. Abitbol, Z.~Ahmed, D.~Barron, R.~B. Thakur, A.~N. Bender, B.~A. Benson,
  C.~A. Bischoff, S.~A. Bryan, J.~E. Carlstrom, C.~L. Chang, D.~T. Chuss, K.~T.
  Crowley, A.~Cukierman, T.~de~Haan, M.~Dobbs, T.~Essinger-Hileman, J.~P.
  Filippini, K.~Ganga, J.~E. Gudmundsson, N.~W. Halverson, S.~Hanany, S.~W.
  Henderson, C.~A. Hill, S.-P.~P. Ho, J.~Hubmayr, K.~Irwin, O.~Jeong, B.~R.
  Johnson, S.~A. Kernasovskiy, J.~M. Kovac, A.~Kusaka, A.~T. Lee, S.~Maria,
  P.~Mauskopf, J.~J. McMahon, L.~Moncelsi, A.~W. Nadolski, J.~M. Nagy, M.~D.
  Niemack, R.~C. O'Brient, S.~Padin, S.~C. Parshley, C.~Pryke, N.~A. Roe,
  K.~Rostem, J.~Ruhl, S.~M. Simon, S.~T. Staggs, A.~Suzuki, E.~R. Switzer,
  O.~Tajima, K.~L. Thompson, P.~Timbie, G.~S. Tucker, J.~D. Vieira, A.~G.
  Vieregg, B.~Westbrook, E.~J. Wollack, K.~W. Yoon, K.~S. Young, and E.~Y.
  Young, \enquote{{CMB-S4 Technology Book, First Edition},}  (2017).

\bibitem{Natoli}
T.~Natoli, B.~Benson, J.~Carlstrom, E.~Chauvin, B.~Clavel, N.~Emerson,
  P.~Gallardo, M.~Niemack, S.~Padin, K.~Schwab \emph{et~al.},
  \enquote{Fabrication of a monolithic 5-meter aluminum reflector for
  millimeter-wavelength observations of the cosmic microwave background,}
  {\protect\JournalTitle{arXiv preprint arXiv:2304.01469}}  (2023).

\bibitem{Nagy_thesis}
J.~M. Nagy, \emph{Optical Development for the SPIDER Balloon-Borne CMB
  Polarimeter} (PhD thesis, Case Western Reserve University, 2017).

\bibitem{Bicep3}
J.~A. {Grayson}, P.~A.~R. {Ade}, Z.~{Ahmed}, K.~D. {Alexander}, M.~{Amiri},
  D.~{Barkats}, S.~J. {Benton}, C.~A. {Bischoff}, J.~J. {Bock}, H.~{Boenish},
  R.~{Bowens-Rubin}, I.~{Buder}, E.~{Bullock}, V.~{Buza}, J.~{Connors}, J.~P.
  {Filippini}, S.~{Fliescher}, M.~{Halpern}, S.~{Harrison}, G.~C. {Hilton},
  V.~V. {Hristov}, H.~{Hui}, K.~D. {Irwin}, J.~{Kang}, K.~S. {Karkare},
  E.~{Karpel}, S.~{Kefeli}, S.~A. {Kernasovskiy}, J.~M. {Kovac}, C.~L. {Kuo},
  E.~M. {Leitch}, M.~{Lueker}, K.~G. {Megerian}, V.~{Monticue}, T.~{Namikawa},
  C.~B. {Netterfield}, H.~T. {Nguyen}, R.~{O'Brient}, R.~W. {Ogburn},
  C.~{Pryke}, C.~D. {Reintsema}, S.~{Richter}, R.~{Schwarz}, C.~{Sorenson},
  C.~D. {Sheehy}, Z.~K. {Staniszewski}, B.~{Steinbach}, G.~P. {Teply}, K.~L.
  {Thompson}, J.~E. {Tolan}, C.~{Tucker}, A.~D. {Turner}, A.~G. {Vieregg},
  A.~{Wandui}, A.~C. {Weber}, D.~V. {Wiebe}, J.~{Willmert}, W.~L.~K. {Wu}, and
  K.~W. {Yoon}, \enquote{{BICEP3 performance overview and planned Keck Array
  upgrade},} in \emph{Millimeter, Submillimeter, and Far-Infrared Detectors and
  Instrumentation for Astronomy VIII,}  vol. 9914 of \emph{Society of
  Photo-Optical Instrumentation Engineers (SPIE) Conference Series} (2016), p.
  99140S.

\bibitem{CLASS}
Z.~{Xu}, M.~K. {Brewer}, P.~F. {Rojas}, Y.~{Li}, K.~{Osumi}, B.~{Pradenas},
  A.~{Ali}, J.~W. {Appel}, C.~L. {Bennett}, R.~{Bustos}, M.~{Chan}, D.~T.
  {Chuss}, J.~{Cleary}, J.~D. {Couto}, S.~{Dahal}, R.~{Datta}, K.~L. {Denis},
  R.~{D{\"u}nner}, J.~R. {Eimer}, T.~{Essinger-Hileman}, D.~{Gothe},
  K.~{Harrington}, J.~{Iuliano}, J.~{Karakla}, T.~A. {Marriage}, N.~J.
  {Miller}, C.~{N{\'u}{\~n}ez}, I.~L. {Padilla}, L.~{Parker}, M.~A. {Petroff},
  R.~{Reeves}, K.~{Rostem}, D.~A. {Nunes Valle}, D.~J. {Watts}, J.~L.
  {Weiland}, E.~J. {Wollack}, and {CLASS Collaboration}, \enquote{{Two-year
  Cosmology Large Angular Scale Surveyor (CLASS) Observations: 40 GHz Telescope
  Pointing, Beam Profile, Window Function, and Polarization Performance},}
  (2020), p. 134.

\bibitem{SPTpol_ref}
J.~E. {Austermann}, K.~A. {Aird}, J.~A. {Beall}, D.~{Becker}, A.~{Bender},
  B.~A. {Benson}, L.~E. {Bleem}, J.~{Britton}, J.~E. {Carlstrom}, C.~L.
  {Chang}, H.~C. {Chiang}, H.~M. {Cho}, T.~M. {Crawford}, A.~T. {Crites},
  A.~{Datesman}, T.~{de Haan}, M.~A. {Dobbs}, E.~M. {George}, N.~W.
  {Halverson}, N.~{Harrington}, J.~W. {Henning}, G.~C. {Hilton}, G.~P.
  {Holder}, W.~L. {Holzapfel}, S.~{Hoover}, N.~{Huang}, J.~{Hubmayr}, K.~D.
  {Irwin}, R.~{Keisler}, J.~{Kennedy}, L.~{Knox}, A.~T. {Lee}, E.~{Leitch},
  D.~{Li}, M.~{Lueker}, D.~P. {Marrone}, J.~J. {McMahon}, J.~{Mehl}, S.~S.
  {Meyer}, T.~E. {Montroy}, T.~{Natoli}, J.~P. {Nibarger}, M.~D. {Niemack},
  V.~{Novosad}, S.~{Padin}, C.~{Pryke}, C.~L. {Reichardt}, J.~E. {Ruhl}, B.~R.
  {Saliwanchik}, J.~T. {Sayre}, K.~K. {Schaffer}, E.~{Shirokoff}, A.~A.
  {Stark}, K.~{Story}, K.~{Vanderlinde}, J.~D. {Vieira}, G.~{Wang},
  R.~{Williamson}, V.~{Yefremenko}, K.~W. {Yoon}, and O.~{Zahn},
  \enquote{{SPTpol: an instrument for CMB polarization measurements with the
  South Pole Telescope},} in \emph{Millimeter, Submillimeter, and Far-Infrared
  Detectors and Instrumentation for Astronomy VI,}  vol. 8452 of \emph{Society
  of Photo-Optical Instrumentation Engineers (SPIE) Conference Series} W.~S.
  {Holland} and J.~{Zmuidzinas}, eds. (2012), p. 84521E.

\bibitem{Healpix}
K.~M. Gorski, E.~Hivon, A.~Banday, B.~D. Wandelt, F.~K. Hansen, M.~Reinecke,
  and M.~Bartelmann, \enquote{Healpix: a framework for high-resolution
  discretization and fast analysis of data distributed on the sphere,}
  {\protect\JournalTitle{The Astrophysical Journal}} \textbf{622}, 759 (2005).

\bibitem{transfer_matrix_paper}
J.~S.~Y. Leung, J.~Hartley, J.~M. Nagy, C.~B. Netterfield, J.~A. Shariff,
  P.~A.~R. Ade, M.~Amiri, S.~J. Benton, A.~S. Bergman, R.~Bihary, J.~J. Bock,
  J.~R. Bond, J.~A. Bonetti, S.~A. Bryan, H.~C. Chiang, C.~R. Contaldi,
  O.~Doré, A.~J. Duivenvoorden, H.~K. Eriksen, M.~Farhang, J.~P. Filippini,
  A.~A. Fraisse, K.~Freese, M.~Galloway, A.~E. Gambrel, N.~N. Gandilo,
  K.~Ganga, R.~Gualtieri, J.~E. Gudmundsson, M.~Halpern, M.~Hasselfield,
  G.~Hilton, W.~Holmes, V.~V. Hristov, Z.~Huang, K.~D. Irwin, W.~C. Jones,
  A.~Karakci, C.~L. Kuo, Z.~D. Kermish, S.~Li, D.~S.~Y. Mak, P.~V. Mason,
  K.~Megerian, L.~Moncelsi, T.~A. Morford, M.~Nolta, R.~O'Brient, B.~Osherson,
  I.~L. Padilla, B.~Racine, A.~S. Rahlin, C.~Reintsema, J.~E. Ruhl, M.~C.
  Runyan, T.~M. Ruud, E.~C. Shaw, C.~Shiu, J.~D. Soler, X.~Song, A.~Trangsrud,
  C.~Tucker, R.~S. Tucker, A.~D. Turner, J.~F. van~der List, A.~C. Weber, I.~K.
  Wehus, S.~Wen, D.~V. Wiebe, and E.~Y. Young, \enquote{A simulation-based
  method for correcting mode coupling in cmb angular power spectra,}
  {\protect\JournalTitle{The Astrophysical Journal}} \textbf{928}, 109 (2022).

\bibitem{Ruze_orig}
J.~Ruze, \enquote{Antenna tolerance theory—a review,}
  {\protect\JournalTitle{Proceedings of the IEEE}} \textbf{54}, 633--640
  (1966).

\bibitem{XLCal}
Q.~{Abarr}, H.~{Awaki}, M.~G. {Baring}, R.~{Bose}, G.~{De Geronimo},
  P.~{Dowkontt}, M.~{Errando}, V.~{Guarino}, K.~{Hattori}, K.~{Hayashida},
  F.~{Imazato}, M.~{Ishida}, N.~K. {Iyer}, F.~{Kislat}, M.~{Kiss},
  T.~{Kitaguchi}, H.~{Krawczynski}, L.~{Lisalda}, H.~{Matake}, Y.~{Maeda},
  H.~{Matsumoto}, T.~{Mineta}, T.~{Miyazawa}, T.~{Mizuno}, T.~{Okajima},
  M.~{Pearce}, B.~F. {Rauch}, F.~{Ryde}, C.~{Shreves}, S.~{Spooner}, T.~A.
  {Stana}, H.~{Takahashi}, M.~{Takeo}, T.~{Tamagawa}, K.~{Tamura},
  H.~{Tsunemi}, N.~{Uchida}, Y.~{Uchida}, A.~T. {West}, E.~A. {Wulf}, and
  R.~{Yamamoto}, \enquote{{XL-Calibur - a second-generation balloon-borne hard
  X-ray polarimetry mission},} {\protect\JournalTitle{Astroparticle Physics}}
  \textbf{126}, 102529 (2021).

\bibitem{PublicDataRepository}
{J. Ruhl}, \enquote{{Public Data Repository},}
  \url{http://github.com/JohnRuhl/public_data} (2023). Online; accessed 21
  February 2023.

\end{thebibliography}

%%%%%%%%%% If preparing manually:
% \newcommand{\enquote}[1]{``#1''}

% \bibitem{Zhang:14}
% Y.~Zhang, S.~Qiao, L.~Sun, Q.~W. Shi, W.~Huang, L.~Li, and Z.~Yang,
%   \enquote{Photoinduced active terahertz metamaterials with nanostructured
%   vanadium dioxide film deposited by sol-gel method,}
%   {\protect\JournalTitle{Optics Express}} \textbf{22}, 11070--11078 (2014).

% {Optical Society}, \enquote{{OSA Publishing},}
%   \url{http://www.osapublishing.org}.

% \bibitem{FORSTER2007}
% P.~Forster, V.~Ramaswamy, P.~Artaxo, T.~Bernsten, R.~Betts, D.~Fahey,
%   J.~Haywood, J.~Lean, D.~Lowe, G.~Myhre, J.~Nganga, R.~Prinn, G.~Raga,
%   M.~Schulz, and R.~V. Dorland, \enquote{Changes in atmospheric consituents and
%   in radiative forcing,} in \enquote{Climate Change 2007: The Physical Science
%   Basis. Contribution of Working Group 1 to the Fourth assesment report of
%   Intergovernmental Panel on Climate Change,}  S.~Solomon, D.~Qin, M.~Manning,
%   Z.~Chen, M.~Marquis, K.~B. Averyt, M.~Tignor, and H.~L. Miler, eds.
%   (Cambridge University Press, 2007).

% \end{thebibliography}

\end{document}